\documentclass[12pt,letterpaper]{article}

\usepackage{graphicx}
\usepackage{amssymb}
\usepackage{graphicx}
\usepackage{subfig}

\usepackage{geometry}
\usepackage{color}
\usepackage{float}
\usepackage{amsmath}
\usepackage{amssymb}

\usepackage{setspace}
\usepackage{newunicodechar}
\usepackage{epstopdf}
\usepackage{algorithm}
\usepackage[colorinlistoftodos]{todonotes}

\usepackage{lineno}

\usepackage{fancyhdr}
\usepackage[T1]{fontenc}
\usepackage[utf8]{inputenc}
\usepackage{authblk}
\usepackage{natbib}

\makeatletter
\renewcommand*{\@fnsymbol}[1]{\ensuremath{\ifcase#1\or \dagger\or*\or  \ddagger\or \mathsection\or \mathparagraph\or \|\or **\or \dagger\dagger \or \ddagger\ddagger \else\@ctrerr\fi}}
\makeatother

\title{Modelling latent individual heterogeneity in mark-recapture data with Dirichlet process priors}
\author[1]{Ford, J.H\thanks{Corresponding author: Jessica.Ford@csiro.au}} 
\author[1]{Patterson, T.A.}
\author[1]{and Bravington, M.V.}
\affil[1]{CSIRO, Castray Esplanade, Hobart, 7001, TAS, Australia.} 

\date{\today}

\begin{document}
	
	\begin{titlepage}
		
		\maketitle

\begin{abstract}
The natural subgroups often seen in mark-recapture studies and the complexity of real mark-recapture data means that parametric and discrete style models can be insufficient. Non-parametric models avoid these often restrictive assumptions. We consider the non-parametric Dirichlet process for modelling latent individual heterogeneity in probability of observation and the probability of remaining in or out of a marine sanctuary. Simulation studies demonstrated accurate estimation of multiple groups of latent individual heterogeneity. Simulations were also used to identify the limits of the Dirichlet process. The ability of the Dirichlet process to pick up unimodal heterogeneity was explored in order to avoid potential spurious multimodality. In application to a subset of the data from the North Atlantic humpback whales we were able to estimate annual population-level variation in usage of the marine sanctuary and three measures of individual-level variation. With the Dirichlet process prior we were able to detect multimodality in each parameter.\\

\textbf{Keywords: } Individual heterogeneity; Dirichlet process prior; hidden Markov model; mark-recapture; North Atlantic humpback whales; marine sanctuary. 

\end{abstract}

\end{titlepage}

\section{Introduction}

Heterogeneity is the rule rather than the exception in nature. Variation between individuals in behaviour, size, physiology or almost any other trait is fundamental to the evolution and dynamics of biological systems \citep{Wilson2010}. Despite this truism, ecological models most often deal with the idealised average individual \citep{Bolnick2003}. While this abstraction from real populations has underpinned much of the key advances in understanding population dynamics, there are instances where it becomes necessary to consider the heterogeneity inherent of ecological systems. 

\medskip{}

A fundamental tool in understanding and assessing the demographics of real populations are mark-recapture studies. These involve following a sample of population members through time to infer abundance and/or survival rates. Individuals are captured, marked and released. At later instances in time, new samples of the population are obtained (e.g. re-captures or re-sightings) \citep{Seber1982}. Recaptures of previously captured individuals can be used to infer survival, and the ratios of recaptures to captures of new individuals can be used in estimating abundance. In this paper we restrict ourselves to the former within a multi-state recapture framework \citep{Lebreton2009}. Here individuals may transit between various latent or partially observed states. These may include states denoting survival status (i.e. dead, alive but unobserved, recaptured dead etc.) or in demographic states (e.g. life stages or age-classes). Such rates of transitions between stages form the basis of estimates used to populate classical population models (e.g. Leslie matrices and similar). 

 \medskip{}
 
Typically, most mark-release recapture modeling treats individuals as homogeneous, conditional on state.  Analysis approaches for mark-release recapture data which have explicitly attempted to account for individual heterogeneity, have most often involved either the use of a pre-set functional form (e.g. a Gaussian), or assignment of individuals to a prespecified number of groups \citep{Pledger2003}. Having to make assumptions about the number of groups \emph{a priori} can result in model selection problems in determining the number of groups \citep{Cubaynes2012}. The use of any pre-set form is limited and limiting, as it enforces strict assumptions on the expected distribution of the population.

 \medskip{}
 
The subgroups often seen in mark-recapture studies and the complexity of real mark-recapture data means that both parametric and discrete style models can be insufficient.  This paper tackles this problem by considering a non-parametric approach, the Dirichlet process prior for modelling latent individual heterogeneity. The Dirichlet process prior is a flexible extension to a parametric model as it avoids assumptions about the functional form of the distribution, and it extends discrete style models to the infinite limit by avoiding any prespecifications about the number of groups \citep{Dorazio2008,Navarro2006}. Despite the appeal of the Dirichlet process prior, it has had little application in mark-recapture analysis perhaps because of its complexity and somewhat confusing literature. One exception is \citet{Dorazio2008} who used the Dirichlet process prior to model animal abundance where heterogeneity in abundance between sites was poorly understood and not directly observable.

\medskip{}

We present a Markov chain Monte Carlo (MCMC) sampler for a hierarchical Bayesian hidden Markov model, applied to mark-recapture data, which allows for individual heterogeneity in both the observation and process components of the model. In doing so we extend the approach presented by \citep{Ford2012} into a fully Bayesian and more flexible approach. The methods we present therefore generalize the existing approaches to mark-recapture and individual heterogeneity data and provide a new set of tools for understanding both accounting for individual heterogeneity in order to derive more robust inferences about populations and also for quantifying the nature and extent of individual heterogeneity in real populations.  

\medskip{}

\subsection{The Dirichlet process prior}
The Dirichlet process was first introduced by \citet{Ferguson1973}. Several well known methods for the representation of a Dirichlet process include the Polya urn scheme \citep{Blackwell1973} or Chinese restaurant process \citep{Pitman2006} and the stick-breaking prior \citep{Sethuraman1994,Ishwaran2001}. Following is a description of the Chinese restaurant process which is the basis of the algorithm used in this paper.  

\medskip{}

\begin{figure}[ht]
\subfloat[]{\includegraphics[height=2in]{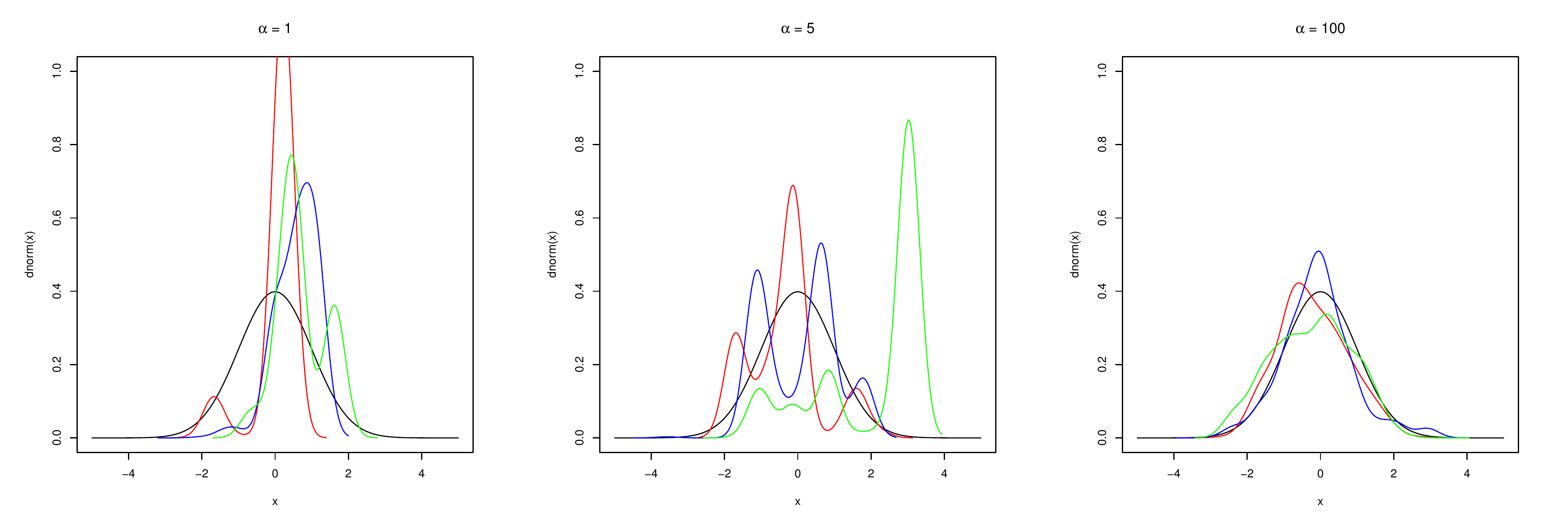}} \\
\subfloat[]{\includegraphics[height=2in]{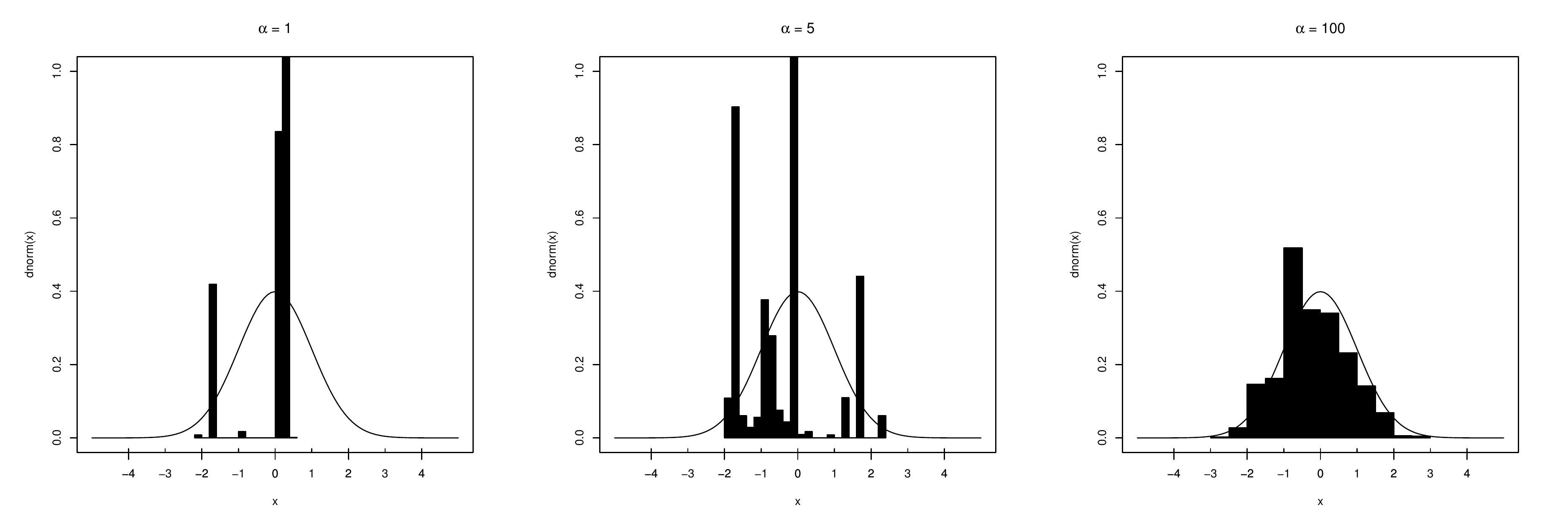}}
\caption{Simulations from a Chinese restaurant process for different weights of $\alpha (1,5,25,100)$, $N=10000$ and $G_{0}\sim N(0,1)$. (a) smoothed density curves for three independent realisations from a Chinese restaurant process for differing values of $\alpha$. The black density line indicates the base distribution $G_{0}$. (b) histograms for the discrete draws that make up one individual realisation.}
\label{fig:Figure1a}
\end{figure}

Consider the analogy of a Chinese restaurant with infinite seating capacity. The first customer enters the restaurant and sits at table one with probability one. Each subsequent customer entering the restaurant chooses a table with probability proportional to the number of people already seated at the table, or a new table proportional to the precision parameter $\alpha$ (this parameter is described in detail below). Customers at the same table are served the same dish; customers at new tables are served a new dish at random. In this sense, individuals at each table receive the same parameter value (dish), and the table number indicates their cluster membership. In general terms, this means that the probability of seeing an already seen cluster is proportional to the number of individuals in that cluster, and the probability of seeing a new cluster is proportional to the precision parameter $\alpha$ (see Figure \ref{fig:Figure1a}).  

\medskip{}

The Dirichlet process is a stochastic process defined as a distribution on distributions and is defined by two quantities: the base distribution $G_{0}$, and the precision parameter $\alpha$. Although the base distribution may be continuous, individual draws $G$ from the Dirichlet process are discrete with probability one \citep{Blackwell1973,Ferguson1973,Neal2000,Sethuraman1994}. This means that draws from a Dirichlet process will be clustered on
a countably infinite set of discrete values. The result is that values will be repeated, as individuals in the same cluster will have the same value. The lower $\alpha$ is, the more variability will be observed between individual realisations, and for any given realisation a small $\alpha$ will correspond to a smaller number of clusters (see Figure \ref{fig:Figure1a}). The influence of $\alpha$ on the number of clusters can be seen in Figure \ref{fig:Figure1a},with the number of clusters increasing with $\alpha$, along with the concentration of draws around $G_{0}$ for large $\alpha$. The number of clusters will tend to $\infty$ with high values of $\alpha$; conversely the number of clusters will tend to $1$ with low values of $\alpha$. In comparison to this non-parametric approach, finite mixtures must specify the number of clusters \emph{a priori}. As such, as $\alpha$ tends to infinity, the Dirichlet process is the limit of the discrete groups approach which assumes a fixed number of groups. In this way $\alpha$ corresponds to the strength of prior belief in the base distribution and the number of groups, or clusters, which are likely to be sampled from it. Note that $G_{0}$ itself will generally be of specified parametric form, e.g. Normal, and will have unknown parameters which are updated separately to the Dirichlet process. 

\medskip{}

A generic Dirichlet process takes the form
\begin{eqnarray*}
y_{i}|\theta_{i} & \sim & F_{i}(\theta_{i})\\
\theta_{i}|G & \sim & G\\
G & \sim & DP(G_{0},\alpha).
\end{eqnarray*}

Here, we assume that data $y_{i}$ are independent conditional on $\theta_{i}$, and $G$ is the mixing distribution over $\theta$ which has Dirichlet process prior $DP(G_{0},\alpha)$.

\medskip{}

MCMC algorithms are the most common approach for inference in Dirichlet processes. \citet{Neal2000} presented several algorithms which use the Chinese restaurant process approach to sample from the posterior distribution of the Dirichlet process. This paper incorporates one of the algorithms developed by \citet{Neal2000}. Alternative samplers include: blocked Gibbs sampling using the stick-breaking representation \citep{Ishwaran2001}; updates using a Metropolis-Hastings framework \citep{Jain2004,Liang2007}; and sequential Monte Carlo \citep{Fearnhead2004}.

\section{Methods}

\subsection{Two-state model used for simulations}
As usual with hidden Markov or multi-state models, the overall model is split into a process part and an observation part. For the process model, we assume that at time $t$ an animal can be in either of two states $S_{it}$: Here and Away, or H/A for short. Changes in the state over time are governed by a Markov process with transition matrix $\gamma$, so (omitting dependence on $i$ for now) for any two states $s$ and $s^{*}$ we have 

\begin{eqnarray*}
\mathbb{P}\left[S_{t+1}=s^{*}\right] & =\sum_{s} & \gamma^{ss*}\mathbb{P}\left[S_{t}=s\right]
\end{eqnarray*}

The four elements of $\gamma$ can be written in terms of just two parameters $\gamma^{HH}$ and $\gamma^{AA}$ (respectively the probabilities of staying Here and staying Away), as follows:

\begin{eqnarray*}
\gamma & = & \begin{pmatrix}\gamma^{HH} & \left(1-\gamma^{HH}\right)\\
\left(1-\gamma^{AA}\right)\, & \gamma^{AA}
\end{pmatrix}
\end{eqnarray*}

For the observation model, there are ``capture attempts" (photo-ID expeditions) at each $t$, in which an animal may be seen if and only if it is Here. Our data for animal $i$ are thus a time series $X{}_{i,t_{1i}:T}$ of $0$ s (not seen) and $1$ s (seen) where $t_{1i}$ denotes the first observation of the animal (see below) and $T$ the most recent expedition. If $X_{it}=1$ then we know 
$S_{it}=\mbox{H}$, but if $X_{it}=0$ the state cannot be determined for certain. Formally, the probability of observation given state is expressed in terms of a parameter $\pi$ by

\begin{eqnarray*}
\mathbb{P}\left[X_{it}=1|S_{it}=s\right] & = & \left\{ \begin{array}{cc}
\pi_{it} & s=\mbox{H}\\
0 & s=\mbox{A}
\end{array}\right.\\
\mathbb{P}\left[X_{it}=0|s\right] & = & 1-\mathbb{P}\left[X_{it}=1|s\right]
\end{eqnarray*}

We start each animal's series at its first sighting of the given year, and condition on $S_{t_{1i}i}=1$. For synthetic data used in this paper, we assume no recruitment and simulate data with all animals present and seen on the first occasion.
 
\subsection{North Atlantic humpback whale data}
The methods developed here are applied to a mark-resight data set on a subpopulation of North Atlantic humpback whales sighted in the Stellwagen Bank National Marine Sanctuary (SBNMS), in the Gulf of Maine. Researchers from the Provincetown Centre for Coastal Studies began documenting North Atlantic humpback whales in the Gulf of Maine in 1975 and have to date individually identified over 1200. Humpback whales (\emph{Megaptera novaeangliae} default) are distributed worldwide, with summer feeding ranges in mid to high-latitudes and winter breeding in low-latitude areas \citep{Clapham1999}. They can be uniquely identified by their natural markings: through the shape of their flukes and through patterns from natural pigmentation \citep{Hammond1986}.
 
\medskip{}

The majority of the North Atlantic humpback whales breed over winter in the West Indies; a small number are thought to use the breeding grounds around the Cape Verde Islands \citep{Stevick1998}.

\medskip{}

During summer, the whales disperse to six summer feeding regions. Although historically treated as a single stock, the six summer feeding regions in the North Atlantic hold relatively discrete subpopulations \citep{Clapham1987}, with individuals demonstrating strong site fidelity to a particular feeding region over many years. Feeding sites include the Gulf of Maine, eastern Canada, west Greenland, Iceland and Norway \citep{Katona1990}, and patterns of movement suggest perhaps four distinct subpopulations \citep{Stevick2006}.

\medskip{}

Individual humpback whales show high maternally directed site fidelity to these summer feeding ranges, as calves follow their mothers from breeding to feeding grounds \citep{Clapham1987}.

\medskip{}

The Gulf of Maine is the southern most summer feeding ground for the North Atlantic humpback whales. Individual humpback whales have been intensively studied in this region since the late 1970s. The SBNMS is one of several important feeding sites for North Atlantic humpback whales which summer in the Gulf of Maine. Due to the consistent aggregation of humpback whales and other marine life, the SBNMS was nominated as a national sanctuary in 1992. This area is not only an important feeding ground for the North Atlantic humpback whales, but is also a busy recreation and transportation area for humans with high levels of commercial and recreational vessel traffic. This overlap has resulted in many injuries to the whales from ship collision and entanglement in fishing gear  \citep{Robbins2004}.
 
\medskip{}

 Although both commercial and recreational fishing are allowed in the sanctuary, regulations have been established which prohibit various other activities such as sand and gravel mining. The sanctuary is a managed resource area equivalent to MPA Category VI \citep{Hoyt2011,IUCN1994}.

\medskip{}

The SBNMS encompasses only a small part of the Gulf of Maine sub population's summer range, and although some individuals are seen regularly there during the summer, none are thought to remain permanently within its boundaries.

\subsection{Three-state model used in application to real data}
The two-state model above is extended to a three-state model for application to real data. A three-state hidden Markov model including death, developed in \citet{Ford2012}, was applied to data from 237 mature North Atlantic humpback whales. In order to draw out a particular instance of heterogeneity in this population we considered only animals seen more than once after the first 8 seasons of data. This is because our primary aim here is to demonstrate heterogeneity within a set of seemingly alike individuals. Younger animals may well be better modeled by the inclusion of age or sex dependent covariates. The three-state model (including death) uses the same implementation as
 the two-state model described above. 
 
\medskip{}

With three states the nine elements of the transition matrix can be written in terms of just three parameters $\gamma^{HH}$, $\gamma^{AA}$, and $\gamma^{D}$ (respectively the probabilities of staying Here, staying Away and Dying in a week) as follows:

\begin{eqnarray*}
\gamma & = & \begin{pmatrix}\gamma^{HH}\left(1-\gamma^{D}\right) & \left(1-\gamma^{HH}\right)\left(1-\gamma^{D}\right)\, & \,\gamma^{D}\\
\left(1-\gamma^{AA}\right)\left(1-\gamma^{D}\right)\, & \gamma^{AA}\left(1-\gamma^{D}\right) & \,\gamma^{D}\\
0 & 0 & \,1
\end{pmatrix}
\end{eqnarray*}

where it is assumed that the probability of death (which is very low relative to the other transition rates) does not depend on whether the animal is Here or Away.

\medskip{}

As sighting effort is focused in the middle of the year, we included all sightings from the 18th week of the year through to the 43rd week. The probability of survival, $P_{surv}$, over the remaining 26 week period was calculated as $P_{surv}=(1-\gamma^{D})^{26}$.

\medskip{}

An extra parameter $q$ was introduced for the probability of being present in the marine sanctuary at the start of the season. We calculated the probability of each state in the first week of the new year to be: 
 \begin{align*}
\mathbb{P}(S_{t}) & =\begin{pmatrix}q*P_{surv}\, & (1-q)*P_{surv}\, & 1-P_{surv}\\
q*P_{surv}\, & (1-q)*P_{surv}\, & 1-P_{surv}\\
0 & 0 & 1
\end{pmatrix}*\mathbb{P}(S_{t-1})
\end{align*}

where $\mathbb{P}(S_{t-1})$ is the vector of state probabilities in the last week of the previous year. 

\subsection{Estimation}
Given a series of observations $X_{1:T}$ and prior distributions on $\pi$  and $\gamma$, our aim is to estimate the posterior distribution using MCMC. The MCMC routine developed in this paper involves four main steps (five in application to real data).

\medskip{}

Individual-level random effects were included on each of $\pi$, $\gamma^{HH}$ and $\gamma^{AA}$ and are updated using the Dirichlet process prior. We assume individual-level parameters to be consistent over time but have allowed for population-level annual variation ($\beta_{yr}$) in probability of remaining Here using logit-links: $\mbox{logit}\gamma_{i,yr}^{HH}=\beta_{yr}+\gamma_{i}^{HH}$. Updates to $\beta_{yr}$, $\gamma^{D}$ (death) and $q$ are assumed to be fixed (not individually variable). 

\medskip{}

One iteration of the MCMC algorithm consists of the following steps: 

\begin{enumerate}
\item Sampling the hidden state chain for all individuals.
\item Calculating summary statistics per individual conditional on its sampled states.
\item Updating the posteriors for individual-level parameters $\pi_{i}$, $\gamma_{i}^{HH}$ and $\gamma_{i}^{AA}$ separately using Gibbs sampling from the Dirichlet process prior.
\item Updating the base distribution and precision parameter:
\begin{enumerate}
\item Updating the base distribution $G_{0}$ using an Independent Metropolis-Hastings sampler with three proposal distributions whose parameters vary across iterations 
\item Updating the precision parameter 
\end{enumerate}
\item Updating population-level fixed effects using an Independent Metropolis-Hastings
 sampler with a fixed proposal distribution: a multivariate t-distribution
 whose mean and variance are set using a preliminary fit from ADMB (see \citep{Ford2012}).
 \end{enumerate}\label{sub:Forward-Backward-recursion}

\subsection{Forward-Backward recursion}
In order to update individual-level parameter values ($\theta$) at each iteration, we require counts of successes and trials for each individual. These counts are obtained from the hidden state chains which are sampled using the Forward-Backward recursion scheme defined by Scott \citeyearpar{Scott2002} and described by Zucchini and MacDonald \citeyearpar{Zucchini2009}. This recursion scheme starts by producing a forward probability vector $\alpha_{2},...,\alpha_{n}$, containing the probabilities of the underlying hidden states for each observation given all observed data up to time $t$. We calculate these forward probabilities, from $1:T$ ($T$ being the length of the observation history), for each state, given the observed data ($X$).  
\begin{eqnarray*}
\alpha_{t}(S_{t}) & = & \mathbb{P}(S_{t}|X_{1:t})\\
 & = & \sum_{S_{t-1}}\mathbb{P}(S_{t-1}|X_{1:t-1})\mathbb{P}(S_{t}|S_{t-1})\mathbb{P}(X_{t}|S_{t})\\
 & = & \sum_{S_{t-1}}\alpha_{t-1}(S{}_{t-1})\mathbb{P}(S_{t}|S_{t-1})\mathbb{P}(X_{t}|S_{t})
\end{eqnarray*}
where $\mathbb{P}(X_{t}|S_{t})$ denotes the probability of the data given the state. Working backwards, we generate a sample path $Z^{(T)}$ of the Markov chain in the order $t=T,T-1,T-2,...,1$, making use of the following proportionality argument: 
\begin{equation}
\mathbb{P}(Z_{t}|x^{(T)},Z_{t+1}^{T},\theta)\propto\alpha_{t}(Z_{t})\mathbb{P}(Z_{t+1}|Z_{t},\theta).\label{eq:sth column-1}
\end{equation}
 The second factor in equation \ref{eq:sth column-1} is simply a one-step transition probability in the Markov chain. 
\medskip{}

\subsection{Counts of successes and trials per individual \label{sub:Counting-of-successes-1}}

Observations for an individual are assumed Binomial with probability $\pi_{i}$. As the Beta prior for $\pi$ is conjugate to the Binomial, the posterior is also Beta. For the probability of observation there is a trial whenever an animal is Here; the outcome is whether it was or wasn't seen. There is no trial when then animal is Away, since it is then guaranteed not to be seen. The counts of successes and trials for the transition probabilities ($\gamma^{HH}$ and $\gamma^{AA}$) are calculated from the sampled state chains. For $\gamma^{HH},$ there is a trial whenever the animal was Here (excluding the final period); the outcome is whether it stayed Here or not. A similar scheme applies to $\gamma^{AA}$. 

\subsection{Gibbs sampling via the Dirichlet process prior.\label{sub:Gibbs-Sampl-viaDPP}}
The individual-level random effects ($\pi_{i}$, $\gamma_{i}^{HH}$ and $\gamma_{i}^{AA}$) are updated separately using a Dirichlet process prior which follows algorithm 8 by \citet{Neal2000} (see algorithm \ref{alg:Algorithm-8-Neal}). 

\begin{algorithm}
Let the state of the Markov chain consist of $c=(c_{1},...,c_{n})$
and $\phi=(\phi_{c}:c\in{c_{1},...,c_{n}})$. Repeatedly sample as
follows:
\medskip{}

\begin{itemize}
\item For $i=1,...,n$ {[}where $n$ indicates the number of individuals{]}:
Let $k^{-}$ be the number of distinct $c_{j}$ for $j\neq i$, and let $h=k^{-}+m$. Label these $c_{j}$ with values in ${1,...,k^{-}}$. If $c_{i}=c_{j}$ for some $j\neq i$, draw values independently from $G_{0}$ for those $\phi_{c}$ for which $k^{-}<c\leq h$. If $c_{i}\neq c_{j}$ for all $j\neq i$, let $c_{i}$ have the label $k^{-}+1$, and draw
values independently from $G_{0}$ for those $\phi_{c}$ for which $k^{-}+1<c\leq h$. Draw a new value for $c_{i}$ from ${1,...,h}$ using the following probabilities: 
\begin{eqnarray*}
P(c_{i}=c|c_{-i},y_{i},\phi_{1},...\phi_{h}) & = & \begin{cases}
b\frac{n_{-i,c}}{n-1+\alpha}\, F(y_{i},\phi_{c})\,\,\mbox{for}\,\,1\leq c\leq k^{-}\\
b\frac{\alpha/m}{n-1+\alpha}\, F(y_{i},\phi_{c})\,\,\mbox{for}\,\, k^{-}\leq c\leq h
\end{cases}
\end{eqnarray*}
where $n_{-i,c}$ is the number of $c_{j}$ for $j\neq i$ that are equal to $c$, and $b$ is the appropriate normalizing constant. Change the state to contain only those $\phi_{c}$ that are now associated with one or more observations. 
\item For all $c\in{c_{1},...,c_{n}}$: Draw a new value from $\phi_{c}\,|\, y_{i}$ such that $c_{i}=c$, or perform some other update to $\phi_{c}$ that leaves this distribution invariant. 
\end{itemize}
\caption{Algorithm 8 by \citet{Neal2000}\label{alg:Algorithm-8-Neal}}
\end{algorithm}

\medskip{}

The algorithms in Neal's paper \citeyearpar{Neal2000} work by assigning individuals to clusters. Due to the clustering property of the Dirichlet process, some of the individual parameter values $\theta_{i}$ will be identical, and each $\theta_{i}$ is associated with a cluster. Indicator variables $c_{i}$ are used to indicate the current cluster membership for each individual (which may change over the course of the MCMC) and the clustering of individuals means that the number of active clusters will typically be much smaller than $N$; $K$ is used to refer to the number of active clusters. For $k=1,...,K$, each cluster $c_{k}$ will have associated parameter value $\phi_{k}$.

\medskip{}

Algorithm 8 in Neal's paper \citeyearpar{Neal2000}, the one implemented here, allows for efficient Gibbs sampling with a non-conjugate distribution. At each iteration, the algorithm temporarily includes $m$ auxiliary components; these are new potential values for clusters, which may or may not actually get individuals assigned to them. For each individual, when updating $c_{i}$, either an existing cluster is chosen or one of these $m$ new components. The probability of joining an existing cluster will be proportional to the number of individuals in that cluster, and the probability of joining a new cluster will be proportional to $\alpha/m$, the prior precision split equally among the auxiliary components. These auxiliary components are generated i.i.d from the base distribution and are discarded at each iteration if not used by the Gibbs sampler (i.e not chosen as a new cluster). The use of auxiliary components avoids the need to integrate with respect to the distribution $G_{0}$ as these auxiliary components represent the new possible components. This approach is similar to methods developed by \citet{MacEachern1998} in that auxiliary components are used to update the model, with the difference that the auxiliary components exist only temporarily in Neal's algorithm. 

\medskip{}

Following Algorithm 8 in Neal's paper \citeyearpar{Neal2000} (see algorithm \ref{alg:Algorithm-8-Neal}), individual parameter values for $\pi_{i}$ , $\gamma_{i}^{HH}$ or $\gamma_{i}^{AA}$ are updated by generating and assigning new clusters. For each class, $c$, the parameter $\phi_{c}$ determines the associated probability for that class; the collection of all $\phi_{c}$ is denoted by $\phi$. In algorithm \ref{alg:Algorithm-8-Neal}, $F(y_{i},\phi_{c})$ is calculated as the density under a Binomial and $c_{i}$ indicates which latent class is associated with observation $y_{i}$, where the numbering of $c_{i}$ is of no significance.

\subsubsection{\textcolor{black}{Updating hyper-parameters ($a,b$) for the base distribution $G_{0}$ and fixed effects\label{sub:Updating-ab-priors}}}
\textcolor{black}{In order to update the hyper-parameters governing the base distribution $G_{0}$ and any population-level fixed effects, it suffices to use the machinery for the Independent Metropolis-Hastings
sampler developed in Ford et al (in submission), which uses a proposal distribution derived from a logit-Normal approximation to the conditional posterior of (a,b). For reference we have included an appendix describing
this method. }

\subsubsection{Updating the precision parameter $\alpha$\label{sub:Updating-alphaprecision}}

Despite its importance, there is a lack of agreement in the literature outlining efficient methods to update the precision parameter \citep{Dorazio2009,Kyung2010,Navarro2006,Escobar1995}. A Gamma($a,b$) prior is commonly used due to its conditional conjugacy property. However, the problem is knowing how to efficiently update the Gamma hyper-parameters ($a,b$). The most recent and concise work in this field is by \citet{Murugiah2012} who propose values for the hyper-parameters which can be used in the presence or absence of information. They suggest that standard use of small $a$ and $b$ can result in high posterior weights for $k=1$ and $k=n$,where $k$ is the number of clusters.  Instead they propose an alternative method which results in $a=b=exp(-0.033n)$, giving a prior mean of unity with increasing standard deviation with larger $n$. The appeal of the method by \citet{Murugiah2012} is that the prior gives less rigid adherence to $G_{0}$ with more data. In cases with small $n$ it will generally be futile to search for, e.g. multimodality, so there is no gain in allowing overly flexible realisations of $G_{0}$. 

\medskip{}

We combine work by \citet{Escobar1995} and \citet{Murugiah2012} to update the precision parameter: methods developed by \citet{Murugiah2012} to update the hyper-parameters are incorporated into the sampling framework developed by \citet{Escobar1995}. \citet{Escobar1995} describe how $\alpha$ can be updated by incorporating an auxiliary variable $\eta$ into the Gamma prior. The formula for updating $\alpha$ is expressed as a mixture of two gamma posteriors, with the conditional mixing parameter for $\alpha$ and $k$, a simple Beta distribution. They found that $p(\alpha|k)$ is the marginal distribution from a joint distribution for $\alpha$ and continuous quantity, $\eta$, such that

\begin{eqnarray*}
p(\alpha,\eta|k) & \propto & p(\alpha)\alpha^{k-1}(\alpha+n)\eta^{\alpha}(1-\eta)^{n-1}
\end{eqnarray*}
where $\eta$ is sampled from a Beta distribution: $(\eta|\alpha,k)\sim B(\alpha+1,n)$.
Taking the conditional posteriors
\begin{eqnarray*}
p(\alpha|\eta,k) & \propto & \alpha^{a+k-2}(\alpha+n)e{}^{-\alpha(b-log(\eta))}\\
 & \propto & \alpha^{a+k-1}e{}^{-\alpha(b-log(\eta))}+n\alpha^{a+k-1}e{}^{-\alpha(b-log(\eta))}
\end{eqnarray*}
when $\alpha>0$ this reduces to a mixture of two Gamma densities \begin{eqnarray*}
(\alpha|\eta,k) & \sim & \pi_{\eta}G(a+k,b-log(\eta))+(1-\pi_{\eta})G(a+k-1,b-log(\eta))
\end{eqnarray*}

where $\pi=\frac{(a_{\alpha}+k-1)}{(n(b_{\alpha}-log(\eta))+a_{\alpha}+k-1)}$. 

\section{Results}
\subsection{Simulation testing\label{sec:Simulation-testing}}
The two-state model was used to test the Dirichlet process using a synthetic data set with 30 individuals, each with 1000 length capture history. We assumed individuals came from (randomly) one of two discrete groups: $\pi$= 0.82 or 0.96; $\gamma^{HH}$= 0.88 or 0.98; $\gamma^{AA}$ = 0.8 or 0.95. Three separate chains were run for 15000 iterations. The chains were arbitrarily thinned to every 2nd update and combined to form one chain of 22500 posterior samples. The chains were thinned to reduce any auto correlation between successive samples \citep{Gilks1996}.
\medskip{}

\begin{figure}
\includegraphics[height=2in]{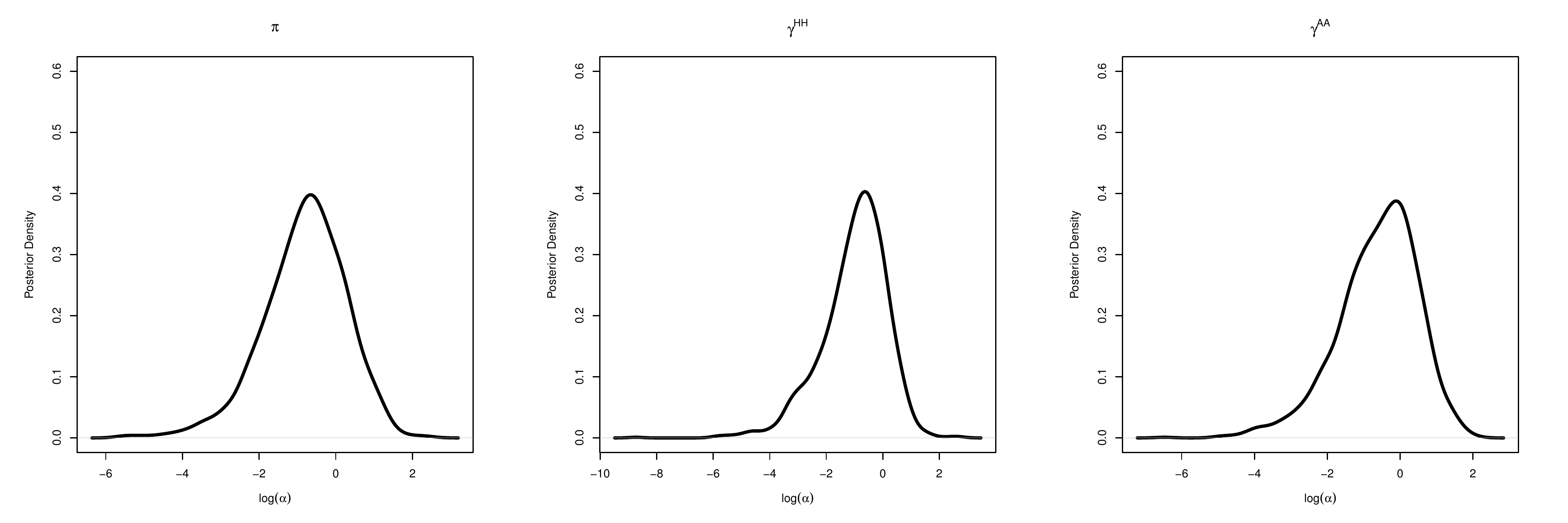}
\caption{Smoothed density curve of posterior estimates for $\mbox{log}(\alpha)$ for each parameter. \label{fig:Figure2}}
\end{figure}

\begin{figure}
\includegraphics[height=2in]{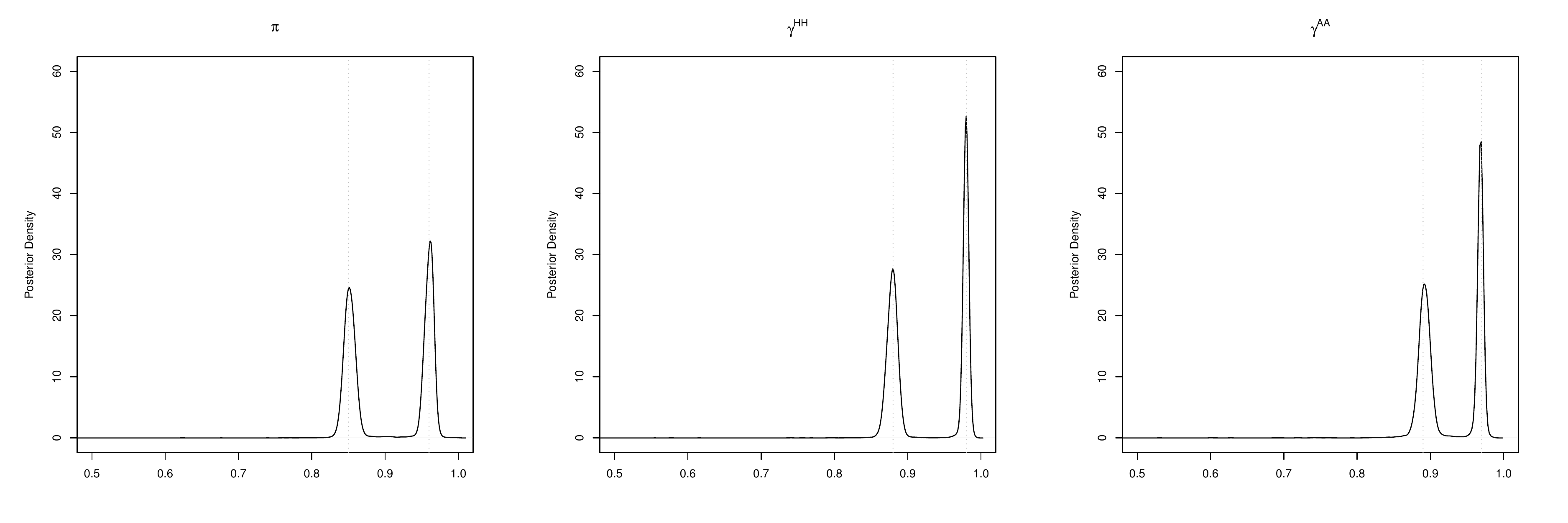}
\caption{Results from 22500 updates combined from three independent chains.Grey dashed vertical lines indicate true value used in data simulation and posterior density of parameters appear to cluster around true values. \label{fig:Figure3}}
\end{figure}

The posterior density of $\mbox{log}(\alpha)$ (Figure \ref{fig:Figure2}) displays standard unimodal form. The posterior distribution of $k$ (the number of clusters) indicated two clusters for each parameter. Figure \ref{fig:Figure3} displays the posterior density for each of $\pi_{i}$ , $\gamma_{i}^{HH}$ and $\gamma_{i}^{AA}$, with posterior estimates for each parameter clustered around the two true values used for simulation. 
\medskip{}

\subsection{Limits of Dirichlet process prior}
The following example is intended to highlight the potential limits of the Dirichlet process in identifying clusters. Data was simulated for 30 animals with 1000 length capture history and run for 15000 iterations, with the first 5000 discarded due to burn-in. Three groups were assumed for both $\pi$$=0.6,0.85,0.96$ and $\gamma^{HH}$$=0.5,0.8,0.95$,
and two groups for $\gamma^{AA}$$=0.89,0.97$. Individuals were randomly assigned to a group for each parameter. 

\medskip{}

\begin{figure}
\includegraphics[height=2in]{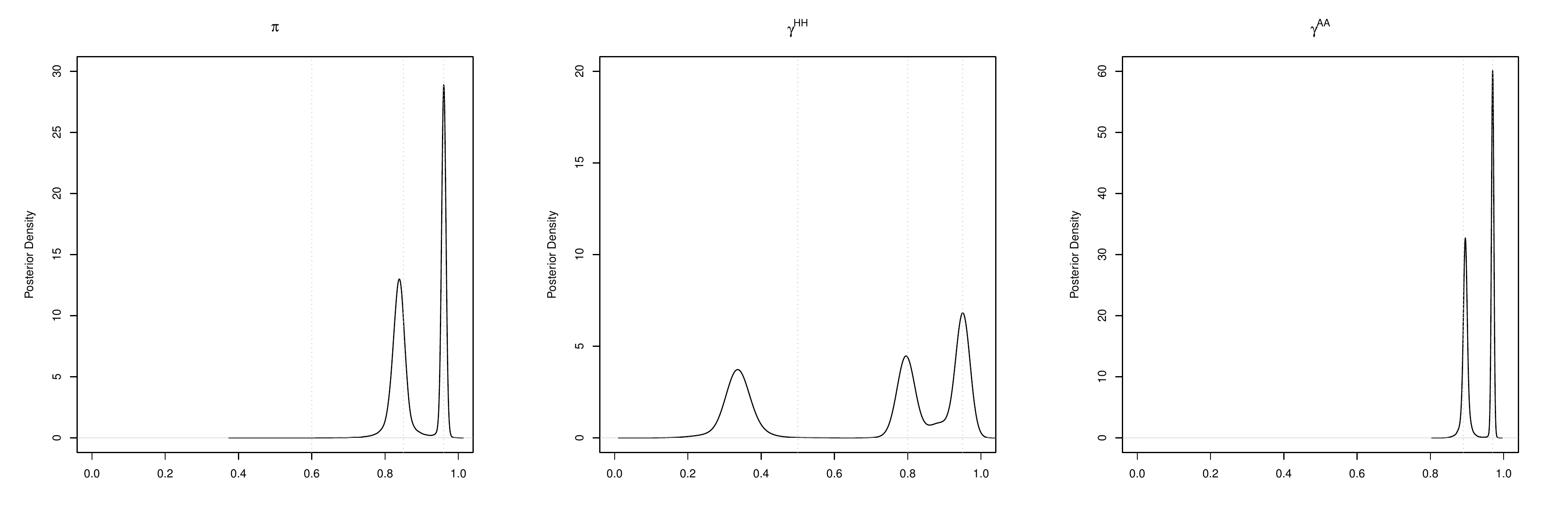}
\caption{Results from 10000 iterations indicating the inability of Dirichlet process prior to distinguish low $\pi$ and low $\gamma^{HH}$. Grey dashed vertical lines indicate the true values used in data simulation. \label{fig:Figure4}}
\end{figure}

Figure \ref{fig:Figure4} indicates the inability of the Dirichlet process prior to distinguish between low $\pi$ and low $\gamma^{HH}$. The results indicate that the lowest true group in $\pi$ ($p=0.6$) could not be distinguished, and that the lowest estimated group in $\gamma^{HH}$  was lower than the actual true values used in data simulation ($\gamma^{HH}=0.5$). At higher probabilities the posterior density of parameters appeared to cluster around the true discrete values used in data simulation. This result is unsurprising due to uninformative data and the resulting inability to distinguish between not being present and not being seen. 

\medskip{}

\subsection{Unimodal distributions}

\medskip{}

\begin{figure}
\includegraphics[height=2in]{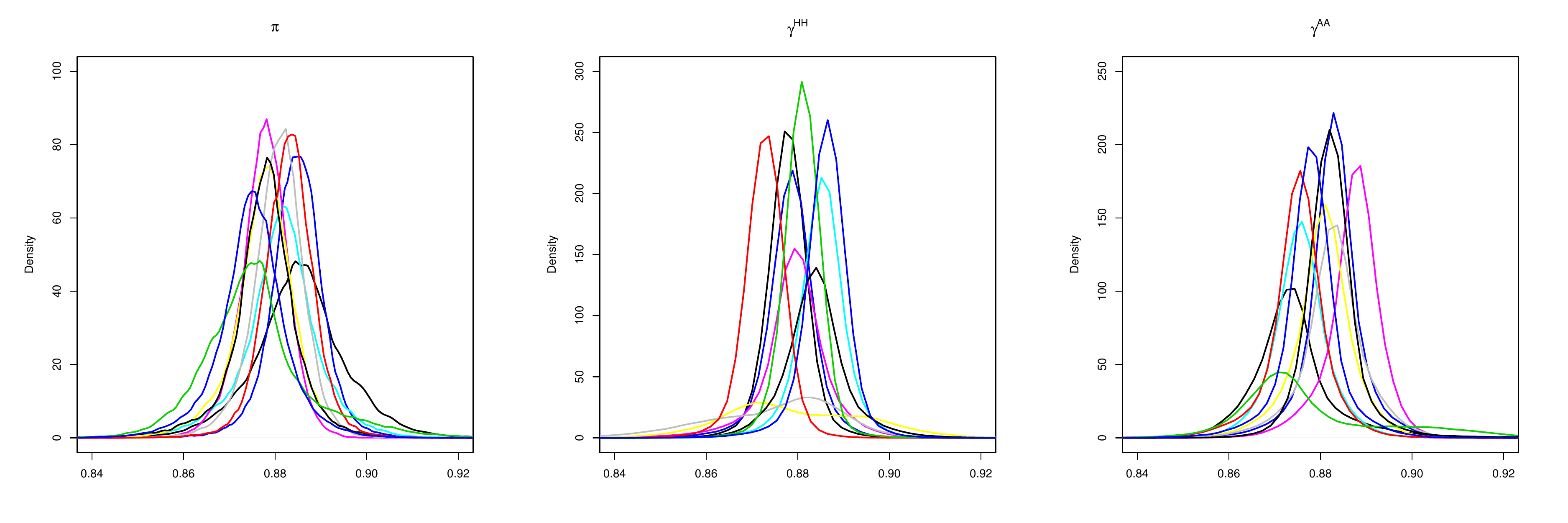}
\caption{Results from 10 independent data sets to test ability of Dirichlet process prior to identify unimodal distribution. Individual parameter values generated using a logit-link and a Normal distribution with low variance. \label{fig:Figure5}}
\end{figure}

One concern with the use of Dirichlet process prior is the potential for spurious multimodality when in fact none is present. To investigate whether this is likely to be a problem we generated 10 synthetic data sets of 30 individuals each with 1000 length capture history. For each parameter ($\pi$, $\gamma^{HH}$ and $\gamma^{AA}$), synthetic data was simulated from a Normal distribution with low variance, $N(2,0.1)$. The MCMC algorithm was run for 10000 iterations. There was no evidence of bi-modality in the results (Figure \ref{fig:Figure5}). The results of this simulation experiment therefore suggest that spurious multimodality given a truly unimodal distribution is unlikely. 

\subsection{North Atlantic humpback whale data analysis\label{sub:WhaleDPPResultssection}}

\begin{figure}
\includegraphics[height=2in]{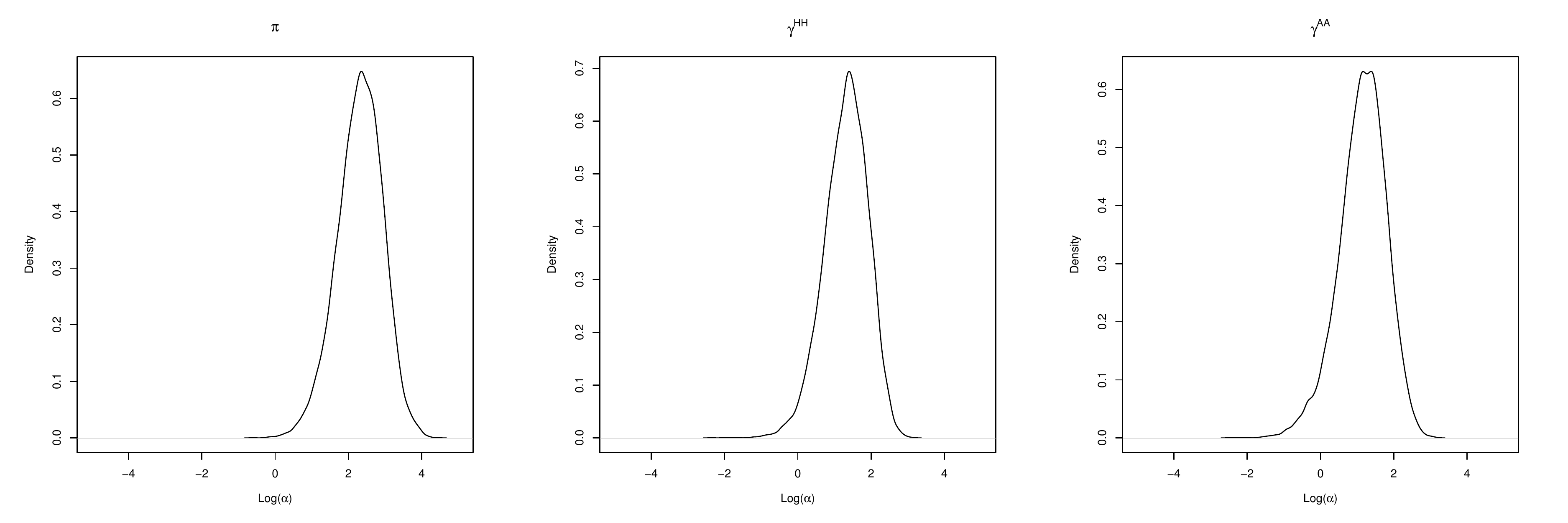}
\caption{Density plot for log($\alpha$), the precision parameter in Dirichlet process prior\label{fig:Figure6}}
\end{figure}

\begin{figure}
\includegraphics[height=2in]{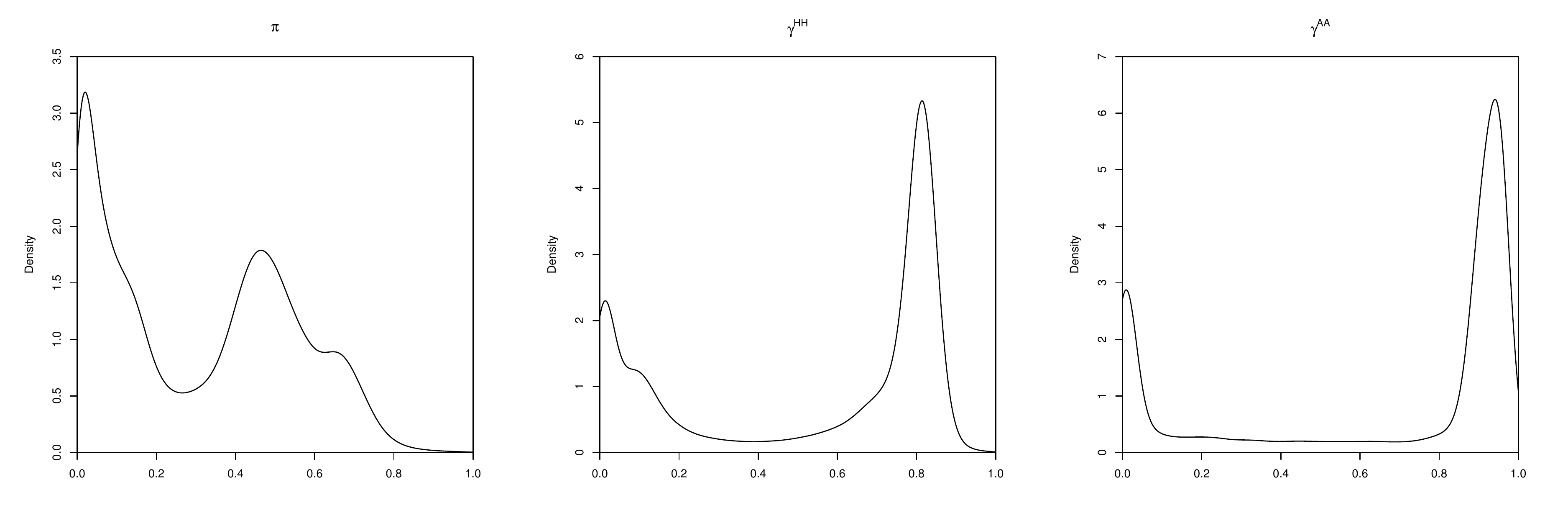}
\caption{Density for $\pi$, $\gamma^{HH}$ and $\gamma^{AA}$ for 25000 iterations for 237 whales. \label{fig:Figure7}}
\end{figure}

One chain was run for 25000 iterations with the first 5000 discarded to burn-in. Density plots for the log of $\alpha$ the precision parameter, indicate expected unimodal density (Figure\ref{fig:Figure6}). The posterior distribution of the number of clusters indicated more variation for $\pi$ compared to both $\gamma^{HH}$ and $\gamma^{AA}$. Figure \ref{fig:Figure7} indicates some multimodality for each of $\pi$, $\gamma^{HH}$ and $\gamma^{AA}$. In each case, low probabilities corresponded to whales seen only a few times. With such an uninformative data history it is difficult to distinguish between not being seen and not being present. As expected, more observations corresponded to higher probability of observation and presence in the marine sanctuary. 

\section{Discussion\label{sec:Discussion}}

In some studies, covariates may adequately explain the majority of individual heterogeneity present in the data. However, in some cases and for certain species (for example cetaceans as considered here), it is unrealistic to expect to be able to collect all necessary covariates, or even to know which covariates would likely explain the heterogeneity. Nonetheless, adequately capturing latent heterogeneity is important to ensure accurate analysis of mark-recapture data. The use of the hidden Markov model combined with the Dirichlet process prior, provides a powerful tool for capturing latent individual heterogeneity. 

\medskip{}

Using simulations, our results show the Dirichlet process prior was able to accurately capture multimodality in three measures of individual heterogeneity: probability of observation, probability of remaining in the marine sanctuary and probability of remaining away. Through simulation studies we were able to explore the accuracy, and limits, of the Dirichlet process prior to distinguish multiple groups using this framework. We found in certain areas of parameter space, the Dirichlet process prior was capable of capturing up to three distinct groups. However, the model was not capable of distinguishing between low probability of observation and low probability of remaining in the marine sanctuary. This aliasing is unsurprising and is due to the lack of information contained in the capture histories \citep{Ford2012}, rather than due to the Dirichlet process prior. 

\medskip{}

\textcolor{black}{In application to North Atlantic humpback whales, we found evidence of multimodality apparent in each parameter. As expected, we found low posterior probabilities corresponded to whales seen only a few times. However, with uninformative data histories identifiability issues are expected as the model cannot discern between 
individuals not being seen or simply not being present. The variation in both the state transition probabilities implies substantial differences in proportion of time spent in the marine }sanctuary\textcolor{black}{. This estimate has implications in the ability to predict the long term usage of the marine }sanctuary\textcolor{black}{{} and for population survival and growth. Whilst this extra uncertainty may have implications for the understanding the population's usage of the marine }sanctuary\textcolor{black}{, it is worth considering what would be inferred from a fixed effect model under similar circumstances. In this case it would be likely to be overestimated \citep{Ford2012} compared to the results from the model here with the Dirichlet process prior. } 

\medskip{}

\textcolor{black}{There are several extensions and applications of the Dirichlet process which were not explored here but are important considerations and interesting areas for future exploration. Additional interesting applications could involve further exploration of correlations between individual random effects: for example, the multiple  behavioural modes indicated that individuals who were often away were more likely to be infrequently observed. With the addition of random effects onto arrival time each year it would be interesting to see the correlation between arrival and departure, and arrival and length of stay in the marine }sanctuary\textcolor{black}{. In future research,  it would be worthwhile investigating the ability of the Dirichlet process to model this, or other, correlations in behaviour. } 

\medskip{}

In comparison to parametric distributions, the Dirichlet process allows for multiple modes in both the observation and state process. Heterogeneity in detection in mark-recapture data has been a hurdle in the accurate estimation of abundance. With the potential to identify multiple modes in the probability of observation, the Dirichlet process has the potential to give more accurate estimates of abundance. The Dirichlet process also has important application to more effective marine spatial planning as it provides a method to more accurately capture the individual behaviour, which translates into more accurate estimations of proportion of time spent in the marine sanctuary. 

\medskip{}

The development of Bayesian hierarchical models has been the focus of much effort in mark-recapture research \citep{King2012}. Despite this, non-parametric approaches have received little attention. We have presented a hierarchical hidden Markov model which allows for both process and observation error and have incorporated the Dirichlet process prior to account for individual heterogeneity on both the observation and process components. We anticipate that this powerful addition to mark-recapture analysis will be useful in application to other problems by allowing for accurate estimation of multiple behavioural modes. 

\medskip{}

\section*{Acknowledgements}
We thank Chris Wilcox and Jooke Robbins for much intellectual input and discussion, and Jooke Robbins and the Provincetown Center for Coastal Studies for data. 

\bibliographystyle{humanbio}
\bibliography{DPP}

\end{document}